\documentclass[twocolumn,nofootinbib,aps,prl,showpacs]{revtex4-1}
\usepackage{graphicx}
\usepackage{float}
\usepackage{amsmath}
\usepackage{amsfonts}
\usepackage{hyperref}
\usepackage{fullpage}
\newcommand{\mathsym}[1]{{}}

\begin{document}
\title{A statistical conservation law in two and three dimensional turbulent flows}
\author{Anna Frishman$^1$, Guido Boffetta$^2$, Filippo De Lillo$^2$, Alex Liberzon$^3$}
\affiliation{$^1$ Physics of Complex Systems, Weizmann Institute of Science, Rehovot 76100 Israel \\
$^2$ Department of Physics and INFN, University of Torino, via P. Giuria 1, 10125 Torino, Italy \\
$^3$ Turbulence Structure Laboratory, School of Mechanical Engineering, Tel Aviv University, Ramat Aviv 69978, Israel }

\date{\today}
\begin{abstract}
Particles in turbulence live complicated lives. It is nonetheless sometimes
possible to find order in this complexity. It was proposed in 
[Falkovich et al., Phys. Rev. Lett. 110, 214502 (2013)] that pairs of Lagrangian tracers at
small scales, in an incompressible isotropic turbulent flow,  have a
statistical conservation law. More specifically, in a d-dimensional flow
the distance $R(t)$ between two neutrally buoyant particles, raised to the
power $-d$ and averaged over velocity realizations, remains at all times equal
to the initial, fixed, separation raised to the same power. In this work we
present evidence from direct numerical simulations of two and three dimensional
turbulence for this conservation. In both cases the conservation is lost when
particles exit the linear flow regime. In 2D we show that, as an extension of the conservation law, a Evans-Cohen-Morriss/Gallavotti-Cohen type fluctuation relation exists.
We also analyse data from a 3D
laboratory experiment [Liberzon et al., Physica D 241, 208 (2012)], finding that although it probes
small scales they are not in the smooth regime. Thus instead of
$\left<R^{-3}\right>$, we look for a similar, power-law-in-separation
conservation law. We show that the existence of an initially slowly varying
function of this form can be predicted but that it does not turn into a
conservation law.
We suggest that the conservation of $\left<R^{-d}\right>$, demonstrated here,
can be used as a check of isotropy,  incompressibility and flow dimensionality
in numerical and laboratory experiments that focus on small scales.
\end{abstract}

\maketitle
\subsection*{Introduction}
The word turbulence is often used as a synonym for turmoil and disorder.
Inherently, particles in a turbulent flow perform an irregular, complicated
motion.
It could therefore come as a surprise that a quantity depending on the
separation between such particles could remain constant during their movement.
Of course, due to the chaotic nature of the flow, one can expect only a
statistical conservation of this type, recovered after averaging over velocity
realizations. 
The subject of the present paper is the verification, on the basis of
both numerical and experimental data of Lagrangian tracers, of the 
conservation law first predicted in \cite{Falkovich2013a}.

Lagrangian conservation laws in turbulence have been studied previously and
have provided much insight into the breaking of scale invariance in such flows
(see \cite{Falkovich2001a} for a review). However, analytical expressions for
them could be derived only for short correlated velocity fields (i.e the
Kraichnan model), and although they were deduced (numerically) and observed in
a Navier-Stokes turbulent flow, \cite{Celani2001}, they were asymptotic laws,
holding only when the initial separation between particles is forgotten.

In \cite{Falkovich2013a} $\left<R(t)^{-d}\right>$ was introduced as an all time
conservation law for a $d$ dimensional flow, where $R(t)=|\textbf{R}(t)|$ is
the time $t$ magnitude of the relative separation between a pair of particles
starting at a fixed distance $\bf R_0$. The conservation is expected for an
isotropic flow and for separations where the velocity difference $u$ scales
linearly with the distance: $u\propto R$, which we will refer to here as a
linear flow. Previously it was believed that it is only an asymptotic in time
conservation law and thus would be difficult to observe \cite{Falkovich2001a} (see also \cite{FalkovichReview} for a historic review).
The result in \cite{Falkovich2013a} opens up the possibility to observe it and
subsequently use it as a check of isotropy and/or incompressibility, as well as flow dimensionality, 
in experiments probing very small scales.
Physically relevant situations for these observations
are phenomena occuring around or below the Kolmogorov scale, such as,
for example, tracer dynamics in cloud physics and Lagrangian statistics
in the direct cascade of two-dimensional turbulence.

The invariance of $\left<R(t)^{-d}\right>$ under the time evolution can be
traced back to a geometrical property of an incompressible linear flow
\cite{Zel'Dovich2006}.
For such a flow, for each velocity realization, the volume of an infinitesimal
d-dimensional hyper-spherical sector, with radius $R_0$ and differential solid
angle $d\Omega_0$, at $t=0$, is equal to its transformation under the flow at
time $t$ \cite{Zel'Dovich2006}:
\begin{equation}
R_0^{d}d\Omega_0=R^{d}(t) d\Omega_t.
\label{eq1}
\end{equation}
Note that it is due to the linearity of the transformation that a spherical
sector is transformed to another spherical sector.
The conservation law directly follows:
for a pair of particles in the flow, starting with the separation vector $\bf R_0$
\begin{equation}
\frac{\left<R(t)^{-d}\right>}{R_0^{-d}}=
\int\frac{\left<R(t)^{-d}\right>}{R_0^{-d} }\frac{d\Omega_0}{S_d}=\int\frac{d\Omega_t}{S_d}=1
\end{equation}
with $S_d$ the volume of the $d-1$-dimensional unit sphere and $d\Omega_0$ parametrizing the direction of $\bf R_0$.
The first equality is a consequence of the assumption that the flow is
statistically isotropic, the average over velocity realizations for a scalar
quantity thus being independent of the direction of $\bf R_0$. In the second
equality the average and the integration are interchanged and equation
(\ref{eq1}) is used.

Recasting the conservation law in the form $\left<e^{-\cal{W}}\right>=1$ with ${\cal W} =\ln(R^d/R_0^d)$ brings to mind the Jarzynski equality \cite{Jarzynski1997,Chetrite2008}, in which ${\cal W}$ is related to entropy production in an out of equilibrium system.
This hints at the possible presence of a Evans-Cohen-Morriss \cite{Evans1993a} or Gallavotti-Cohen \cite{Gallavotti1995} type fluctuation relation. Below we demonstrate that such an extension of the conservation law is indeed possible for symplectic flows. In particular, since a (linearised) two dimensional incompressible flow is already symplectic, the existence of a fluctuation relation is guaranteed and no further assumption, such as time reversibility, is required. 
Resemblance to the Jarzynski equality also provides a different perspective on the latter- it can be thought of as a statistical conservation law.   

In the following we provide a direct confirmation of conservation of
$\left<R(t)^{-d}\right>$ in an isotropic fully developed turbulent flow. We
first present the results from a direct numerical simulation of a 2D flow, with
large scale forcing and friction. For scales much smaller than the forcing, in
the direct cascade, the flow is linear and the theory applies. We show that as
long as the separation between particles remains in this regime,
$\left<R(t)^{-2}\right>$ remains constant as well.
In addition, we find the symmetry relation
$\left<\left(R(t)/R_0\right)^q\right>=\left<\left(R(t)/R_0\right)^{\tilde{q}}\right>$
with $\tilde{q}=-q-2$ as a generalization of the conservation of
$\left<R(t)^{-2}\right>$. At long times this yields a
Gallavotti-Cohen type fluctuation relation as discussed above. 

Next we consider data from a direct numerical simulation of $3$-dimensional
turbulence and study particles with initial separations in the dissipative
range, where approximately $\left<\textbf{u}^2\right> \propto R^2$. We observe the
conservation of $\left<R(t)^{-3}\right>$ ending in a regime where
$\left<R(t)^{-3}\right>$ no longer converges. At those times the separations
distribution at small scales is induced by the dynamics in the inertial range
rather than the dissipative range of scales.

Finally, we turn to data of $3D$ turbulence from a laboratory experiment.
Initial particle separations are taken at scales comparable with the 
Kolmogorv scale of the flow.
We find that, in distinction from the numerical simulation, the scaling of
the velocity is far from the approximation $\left<\textbf{u}^2\right> \propto R^2$ for
such separations. Also, deviations from an isotropic flow are observed. Thus,
we check instead if an analogue of $\left<R^{-3}\right>$, in the form of a
power law, exists for the scales available in the experiment. We show that a
slowly varying function of this form can be predicted using the statistics of
the pairs relative velocity and acceleration at $t=0$. However, it does not
appear to be a true conservation law.

\subsection*{Conservation in 2D -DNS results}
The motion of Lagrangian tracers is numerically integrated
in the direct cascade of two-dimensional turbulence with friction, in a square
box with side length $L=2\pi$. The flow is generated by a large scale,
$\delta$-correlated, random forcing at a wave-number corresponding to the box
scale which injects enstrophy at a rate $\theta_I$.
The parameters of the simulation are taken from \cite{Boffetta2002a}.
The linear friction is sufficiently strong to generate a
velocity field with energy spectrum exponent close to $~4.5$ and a power-law
decaying enstrophy flux, i.e no logarithmic corrections to the leading order scaling
are to be expected \cite{Bernard2000},\cite{Boffetta2012a}.
The viscous enstrophy dissipation rate $\theta_{\nu}$, together with the kinematic
viscosity $\nu$, define the smallest dissipative scale
$\eta \simeq \nu^{1/2} \theta_{\nu}^{-1/6}$ \cite{Boffetta2012a} which is
used as a reference scale. Time is made dimensionless with the vorticity
characteristic time $\tau_{\eta}=(2 Z)^{-1/2}$ ($Z$ represents the mean enstrophy)
and consequently the reference velocity is $u_{\eta}=\eta/\tau_{\eta}$.

\begin{figure}
    \includegraphics[width=1\linewidth]{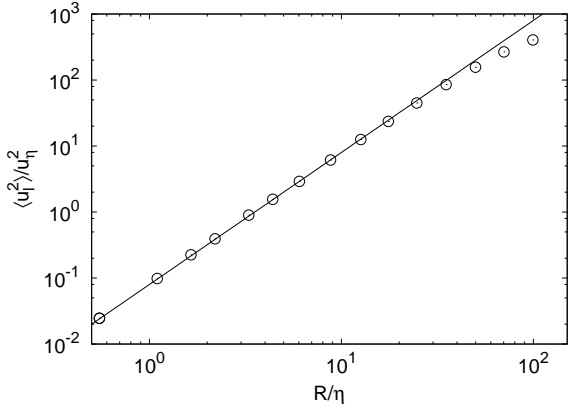}
    \centering
    \caption{Second order longitudinal structure function of velocity in the
2D direct cascade simulation. The line represents the behaviour
$\left<u_l^2(R)\right>\propto R^2$ }
    \label{fig:1}
\end{figure}

Fig. \ref{fig:1} shows the longitudinal structure function of the velocity,
where we denote by $u_l=\textbf{u}\cdot \hat{R}$ the longitudinal velocity
difference at scale $R$. A scaling consistent with a linear flow behaviour is
observed for $R \lesssim 10 \eta$.

In stationary conditions, $N_p=16384$ particle pairs are introduced into the
flow with homogeneous distribution and initial separation $R_0$ and their
trajectories are evolved in time.
The moments of separation $\left<R(t)^q\right>$ are computed by averaging over
$N_p$ and $N_{run}=100$ independent runs.

%%%%%%%%%%%%%%%%%%%%%%%%%%%%%%%%%%%%%%%%%%%%%%%%%%%%%%%
%%%%%%%%%%%%%%%%%%%%%%%%%%%%%%%%%%%%%%%%%%%%%%%%%%%%%%%

\begin{figure}
 \includegraphics[width=1\linewidth]{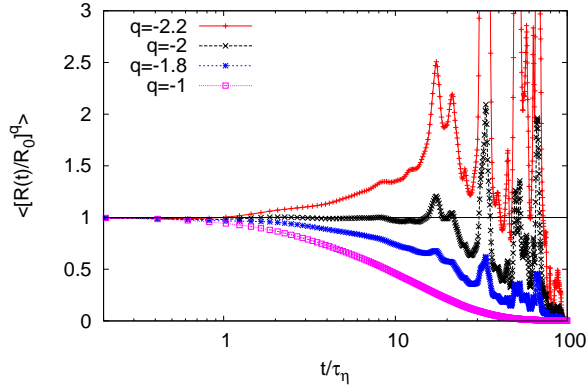}
 \centering
 \caption{Time evolution of different moments of relative separation $R(t)$ as
a function of time in the 2D simulations ($R_0=1.1\eta$). Different point styles represent
$q=-1.0$, $q=-1.8$, $q=-2.0$ and $q=-2.2$ (from bottom to top).}
 \label{fig:2}
\end{figure}

In Fig. \ref{fig:2} we present the time evolution of $\left<R^q(t)\right>$
for $q=-2.2,-2,-1.8,-1$. While the moment for $q=-2.2$ ($q=-1.8$) grows (decays), doing so exponentially for $t>2\tau_{\eta}$, the moment with $q=-2=-d$ is conserved up to a time $t \approx
10 \tau_{\eta}$. 
At this time the average particle separation reaches $R \sim 10\eta$ where
$\left<u_l^2\right>$ in Fig. \ref{fig:1} deviates from a linear flow. We observe that the exponential growth of $\left<R\right>$ is inhibited at
approximately the same time (not shown). On the other hand, as can be seen in Fig. \ref{fig:2}, the exponential decrease of
$\left<R^{-1}\right>$, which is less sensitive to pairs with large
separations, lasts throughout the observation time.

%%%%%%%%%%%%%%%%%%%%%%%%%%%%%%%%%%%%%%%%%%%%%%%%%%%%%%%
%%%%%%%%%%%%%%%%%%%%%%%%%%%%%%%%%%%%%%%%%%%%%%%%%%%%%%%

The exponential time dependence of $\left<R^q(t)\right>$ observed in Fig. 
\ref{fig:2} is expected to be a long time feature of any linear flow with
temporal correlations decaying fast enough \cite{Falkovich2001a},
\cite{Balkovsky1999a}. Specifically, $E(q,t)\equiv\ln
\left<\left(R(t)/R_0\right)^q\right>$, the cumulant generating function of
$\beta= \ln \left(R(t)/R_0\right)$, should take at long times the form
$E(q,t)=L(q)t$. We remark that the finite-time Lyapunov exponent $\sigma=\beta/t$ tends to
the Lagrangian Lyapunov exponent, $\lambda$, in the long time limit \cite{cencini2010chaos}. 

%%%%%%%%%%%%%%%%%%%%%%%%%%%%%%%%%%%%%%%%%%%%%%%%
\begin{figure}
    \includegraphics[width=1\linewidth]{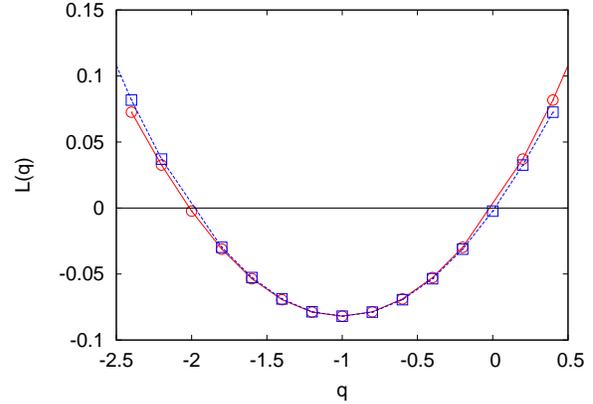}
    \centering
    \caption{$L(q)$ obtained from fitting $\left<R(t)^q\right>$ with 
$R_0^q e^{L(q)t}$ in the time interval $2\le t/\tau_{\eta} \le 10$
for the 2D simulations ($R_0=1.1\eta$).
In order to show the symmetry with respect to $q=-1$, both $L(q)$ (red circles)
and $L(-2-q)$ (blue squares) are plotted.}
    \label{fig:3}
\end{figure}
\begin{figure}[h!]
\centering
\includegraphics[width=\linewidth]{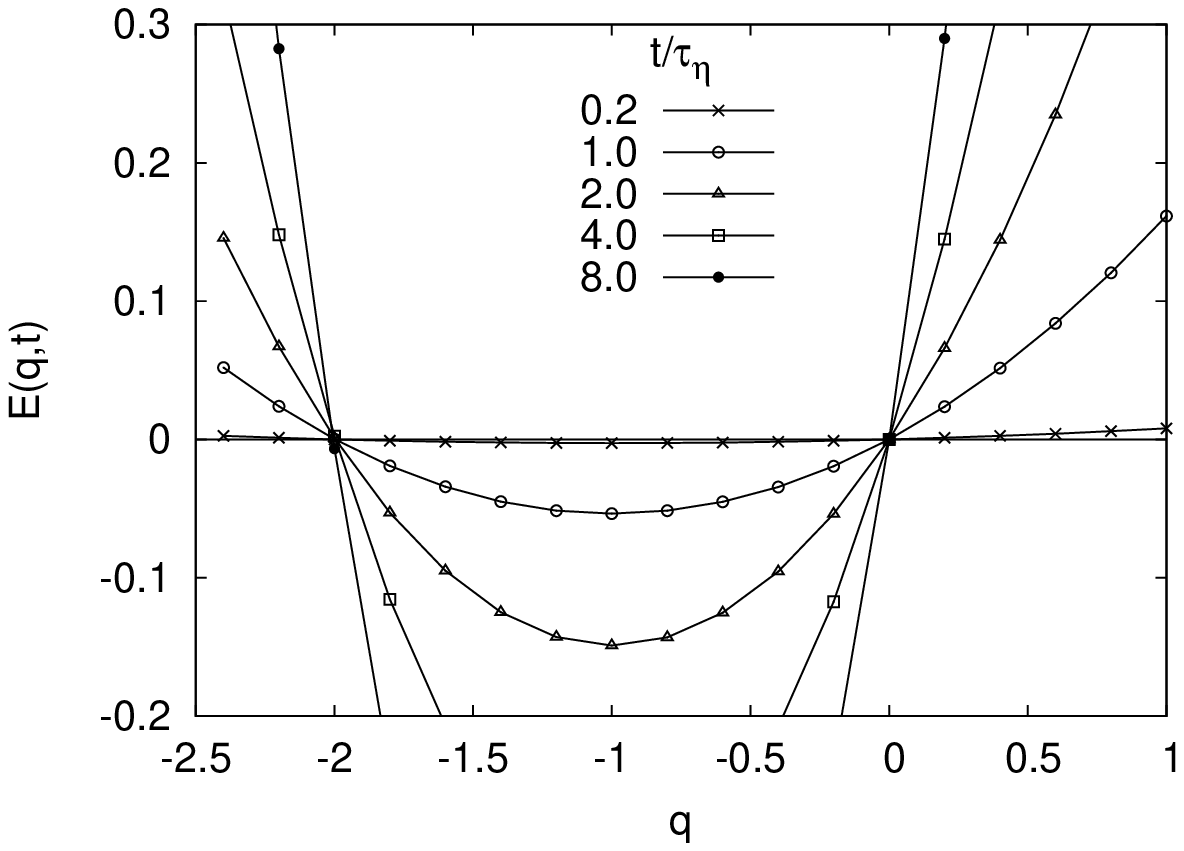}
\centering
\includegraphics[width=\linewidth]{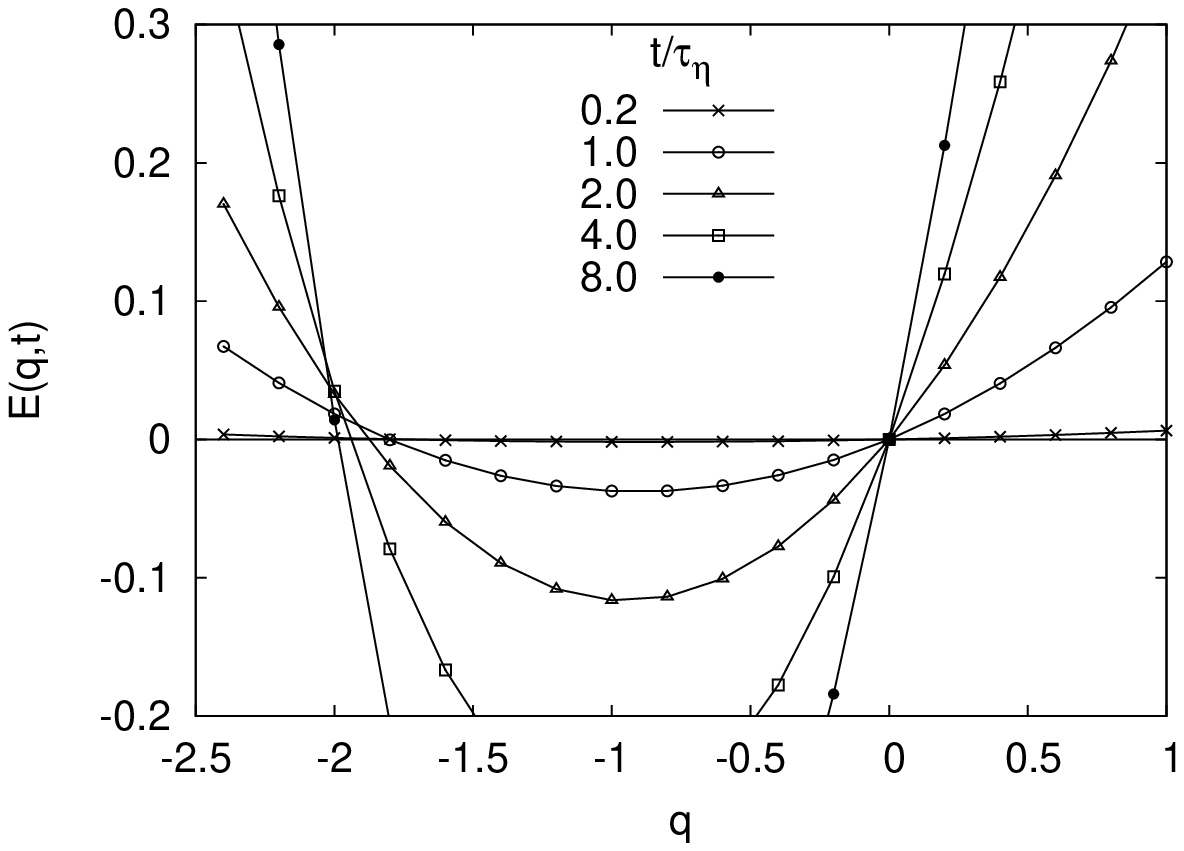}
\centering
\includegraphics[width=\linewidth]{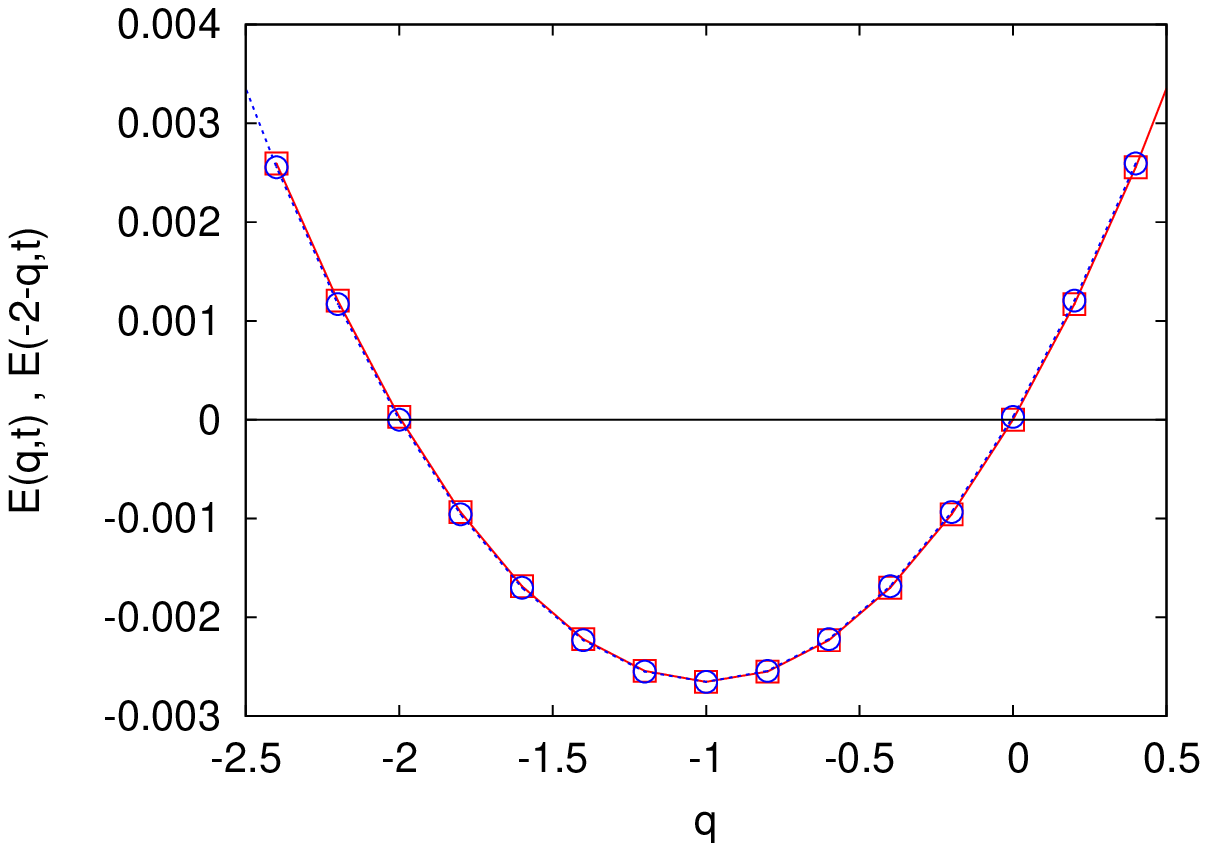}
\caption{$\ln\left<\left(R(t)/ R_0\right)^q\right>$ as a function of $q$
for two different initial separations $R_0=1.1\eta$ (upper plot) and 
$R_0=50\eta$ (middle plot). Different curves represent different times.
Lower plot:
$E(q,t)$ (red circles) and $E(-2-q,t)$ (blue squares)
for $t=0.2 \tau_{\eta}$ and $R_0=1.1 \eta$.
Data from 2D simulations.}
\label{fig:5}
\end{figure}

%%%%%%%%%%%%%%%%%%%%%%%%%%%%%%%%%%%%%%%%%%%%%%%%
We obtain $L(q)$ shown in Fig. \ref{fig:3} by fitting $\left<R(t)^q\right>=R_0^q e^{L(q)t}$ for times $0.4\le \lambda t\le 2$. Evidently $L(q)$ is perfectly symmetric with respect to $q=-d/2=-1$, i.e $L(q)=L(-q-2)$, implying in particular that $L(-2)=0$ as expected.
This symmetry can be extended to the symmetry $E(q,t)=E(-q-2,t)$ holding at any time $t$, as we demonstrate in the bottom figure of Fig. \ref{fig:5} for $t=0.2\tau_{\eta}$. This is a general property of a two-dimensional linear incompressible and isotropic flow.
In fact, this follows from the relation
\begin{equation}
\label{eq2}
\int \left(\frac{R(t)}{R_0}\right)^{q} d\Omega_0=\int \left(\frac{R(t)}{R_0}\right)^{-q-2} d\Omega_0
\end{equation}
which, as we will show, holds for every velocity realization. Indeed, using the definition of $E(q)$ one obtains $E(q,t)=E(-q-2,t)$ for an isotropic flow by taking the average of (\ref{eq2}) over velocity realizations, conditioned that the averages exist.

To demonstrate (\ref{eq2}) we recall that for a linear flow with a given velocity realization the following decomposition can be used
\begin{equation}
\label{dec}
\frac{R(t)^2}{R_0^2}=\hat{R}_0^{T} O^T(t)\Lambda(t)O(t)\hat{R}_0
\end{equation}
where $O(t)$ is an orthogonal matrix and $\Lambda$ a diagonal matrix with entries $e^{\rho_1(t)}$ and $e^{\rho_2(t)}$  \cite{cardy2008non}\cite{Falkovich2001a}.
For an incompressible flow $\det \Lambda=1$ so that in 2D $\rho_1=-\rho_2\equiv \rho$.
Using (\ref{dec}) the integration over $\hat{R}_0$ can be written as
\begin{equation}
\label{Rn0}
\begin{split}
\int \left(\frac{R(t)}{R_0}\right)^q d\Omega_0= \int_0^{2\pi} \left[\hat{e}^{T}(\theta)\Lambda(t)\hat{e}(\theta)\right]^{q/2}d\theta
\end{split}
\end{equation}
where $\hat{e}=(\cos(\theta),\sin(\theta))$ and a change of integration variable absorbed
the additional rotation $O(t)$ in (\ref{dec}).
On the other hand, (\ref{eq1}) implies
\begin{equation}
\label{dyn}
\int \left(\frac{R(t)}{R_0}\right)^q d\Omega_0=\int \left(\frac{R_0}{R(t)}\right)^{-q-2} d\Omega_t
\end{equation}
and $R_0/R(t)$ can be decomposed, similarly to (\ref{dec}), using the flow backward in time
\begin{equation}
\label{par2}
\frac{R_0^2}{R(t)^2}=\hat{R}_t^{T} O^T(-t)\Lambda(-t)O(-t)\hat{R}_t.
\end{equation}
The inversion of the linear transformation between $\textbf{R}(t)$ and $\textbf{R}(0)$
implies $\Lambda(-t)=\Lambda^{-1}(t)$ \cite{cardy2008non}. Then, $\Lambda^{-1}(t)$ is related to $\Lambda(t)$ by conjugation with a rotation matrix since $\Lambda=diag(e^{\rho},e^{-\rho})$. This property is actually a consequence of the symplectic structure of a 2D incompressible flow- for a symplectic flow the eigenvalues of $\Lambda$ come in pairs of the form $e^{\pm\rho_i}$ \cite{cencini2010chaos}. \footnote{
As this is the basic symmetry leading to the relations (\ref{Pt}) and (\ref{G}), these relations, with the replacement of $2$ by $2N$, would also hold for a linear random/chaotic N-dimensional Hamiltonian flow that is statistically isotropic, $R$ denoting the separation between two points in phase space.} Therefore for any $m$, shifting the integration variable similarly to (\ref{Rn0}),
\begin{equation}
\begin{split}
\int \left(\frac{R_0}{R(t)}\right)^{m} d\Omega_t=  \int_0^{2\pi}\left[\hat{e}^{T}(\theta) \Lambda(t)\hat{e}(\theta)\right]^{m/2}d\theta \nonumber
\end{split}
\end{equation}
which by comparison to (\ref{Rn0}) leads to
\begin{equation}
\label{Rnt}
\begin{split}
\int \left(\frac{R_0}{R(t)}\right)^{m} d\Omega_t= \int \left( \frac{R_0}{R(t)}\right)^{-m} d\Omega_0.
\end{split}
\end{equation}
Finally, combining equation (\ref{Rnt}) with $m=-q-2$ together with equation (\ref{dyn}) yields (\ref{eq2}).

The symmetry of $E(q)$ implies a symmetry of the probability density function (pdf) $P\left(\beta(t)\right)$, $\beta= \ln \left(R(t)/R_0\right)$, which can be obtained by first extending\footnote{This can be done since the considerations above did not depend on $q$ being real, only on convergence of the integrals. It seems reasonable that if they converge for real $q$ they would also converge for its complex values.} the symmetry of $E(q)$ to complex $q$ and then retrieving the pdf by an inverse Laplace transform of $\exp[E(q)]$. Using the symmetry and a change of variables in the integration, one gets for any time $t$
\begin{equation}
\label{Pt}
\frac{P(\beta(t)=y)}{P(\beta(t)=-y)}=e^{2 y }= J(y)
\end{equation}
where $J(\ln \left(R(t)/R_0\right))$ is the Jacobian of the transformation (\ref{eq1}) from time zero to time $t$, in 2D.

%%%%%%%%%%%%%%%%%%%%%%%%%%%%%%%%%%%%%%%%%%%%%%%%
\begin{figure}    \includegraphics[width=1\linewidth]{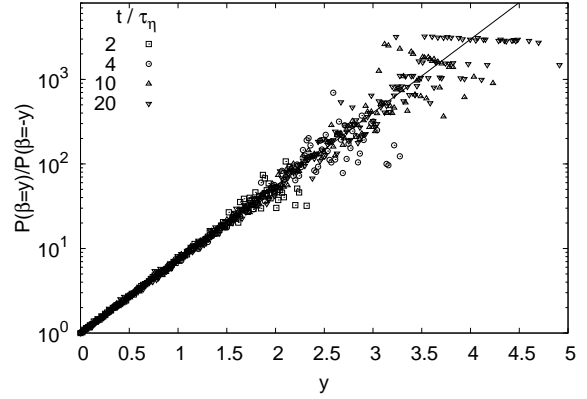}    \centering    \caption{Verification of the symmetry relation (\ref{Pt}), $\beta= \ln \left(R(t)/R_0\right)$, in the 2D simulations. The solid straight line corresponds to the analytical prediction.}
    \label{fig:4}
\end{figure}
%%%%%%%%%%%%%%%%%%%%%%%%%%%%%%%%%%%%%%%%%%%%%%%%

In the asymptotically long-time limit (\ref{Pt}), which is numerically verified in Fig. \ref{fig:4}, turns into an Evans-Cohen-Morriss \cite{Evans1993a} or Gallavotti-Cohen \cite{Gallavotti1995} type fluctuation relation.
Indeed, the function $L(q)$ is the Legendre transform of the long-time large deviation function $G(\sigma)$ so that the symmetry in $L(q)$ implies:
\begin{equation}
\label{G}
G(-\sigma)=G(\sigma)+2\sigma
\end{equation}
which can also be deduced directly from (\ref{Pt}).
This relation resembles the Evans-Cohen-Morriss-Gallavotti-Cohen relation, and even more so the relation found in \cite{Chetrite2007b}, but is different from them. Note that we do not assume a time reversible velocity ensemble and work with an incompressible flow (implying zero entropy production in the context of dynamical systems). The two dimensionality of the flow plays a crucial role in the derivation.

In Fig. \ref{fig:5} we show $E(q,t)$ for short times, $t \le 8 \tau_{\eta}$
as a function of $q$, comparing two distinct initial separations.
For $R_0=1.1\eta$ all curves cross zero at $q=-2$, up to time
$t=8\tau_{\eta}$, while for $R_0=50\eta$, which is at the
border of the linear scaling range, the initial crossing is shifted
above $-2$, and the intersection point changes with time (see middle Figure).
The appearance of the initial crossing above $-2$ in the latter can be explained using a Taylor
expansion of $\left<\left(R(t)/ R_0\right)^q\right>$ around $t=0$, see
equations (\ref{Rnstar}),(\ref{nstariso}) and (\ref{zeta2}) in the analysis of
the experimental results, as well as \cite{Falkovich2013a}.  

%%%%%%%%%%%%%%%%%%%%%%%%%%%%%%%%%%%%%%%%%%%%%%%%
%%%%%%%%%%%%%%%%%%%%%%%%%%%%%%%%%%%%%%%%%%%%%%%%

%%%%%%%%%%%%%%%%%%%%%%%%%%%%%%%%%%%%%%%%%%%%%%%%%%%%%%%%%%%%%%%%%%%%%
\subsection*{Conservation in 3D - DNS results}
Lagrangian tracers are introduced into a numerical simulation of
three-dimensional turbulence in a cubic box at $Re_{\lambda}=107$. The tracers
are placed in the flow when it has reached a steady state and their
trajectories are numerically integrated.
Turbulence is generated by a large scale, $\delta$-correlated random forcing.
Small scales are well resolved in the simulation 
(for which  $k_{max} \eta= 1.3$). More than 130000 pairs were integrated for times up to about 10 $\tau_\eta$ for 300 realizations each, while more than 30000 pairs were integrated for 150 realizations for longer times. The two datasets are overlapping for shorter times so that consistency of the statistics could be checked.

\begin{figure}
%%%%%%%%%%%%%%%%%%%%%%%%%%%%%%%%%%%%%%%%%%%%%%%%
 \includegraphics[width=1\linewidth]{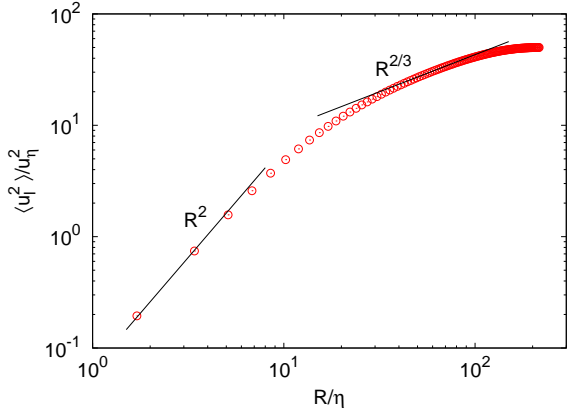}
 \centering
 \caption{ Second order longitudinal velocity structure function for the 3D
numerical simulation. The two lines represent the smooth $R^2$ and rough
$R^{2/3}$ behaviors.}
\label{fig:6}
\end{figure}

In Fig. \ref{fig:6} we show the second order longitudinal velocity structure function. Up to separations of about $7\eta$ the scaling exponent of $\left<u_l^2(R)\right>$ is close to $2$, although deviations are also observed earlier.
%%%%%%%%%%%%%%%%%%%%%%%%%%%%%%%%%%%%%%%%%%%%%%%%

\begin{figure}[h!]
%\begin{subfigure}{0.7\linewidth}
  \centering
  \includegraphics[width=\linewidth]{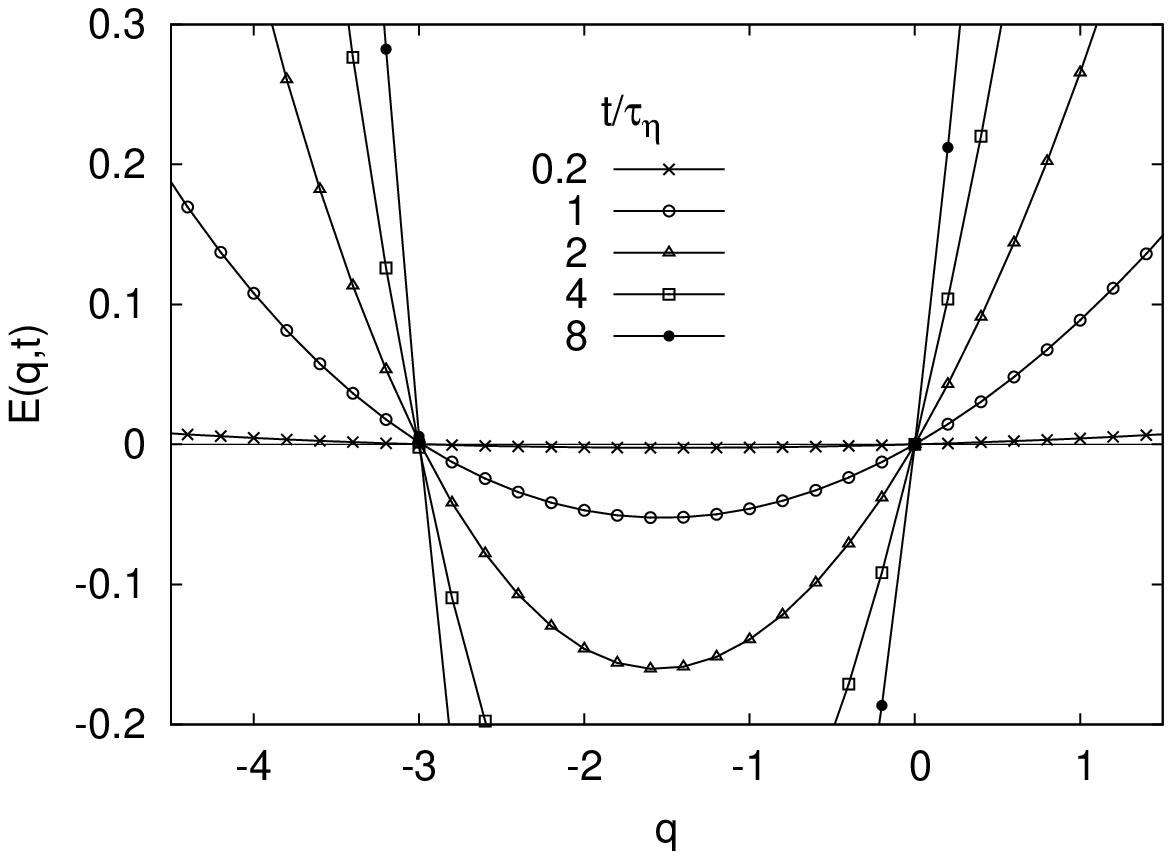}
%  \caption{$R_0=1.7\eta$}
%  \label{a}
%\end{subfigure} \\
%\begin{subfigure}{.5\linewidth}
  \centering
  \includegraphics[width=\linewidth]{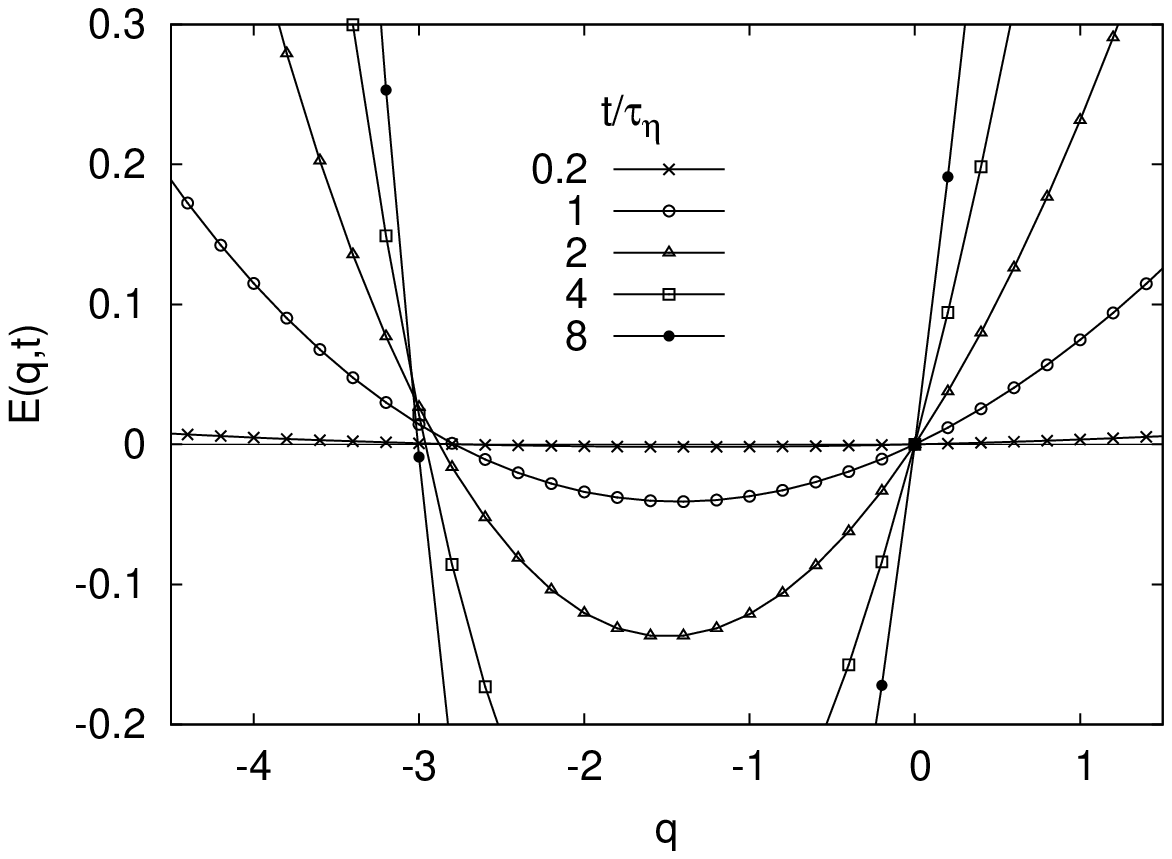}
%  \caption{$R_0=6.76\eta$}
%  \label{b}
%\end{subfigure}%
%\begin{subfigure}{.5\linewidth}
  \centering
  \includegraphics[width=\linewidth]{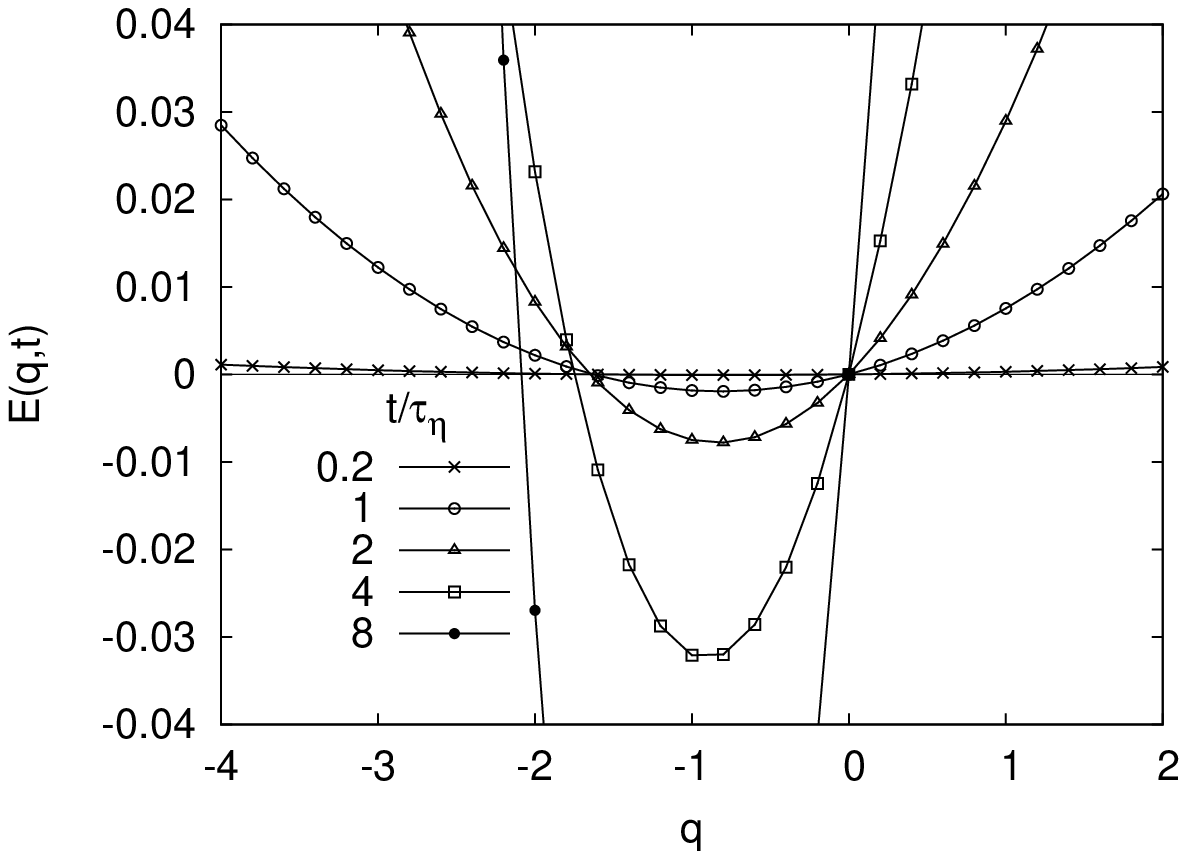}
%  \caption{$R_0=70\eta$}
%  \label{c}
%\end{subfigure}
\caption{$\ln\left<\left(R(t)/ R_0\right)^q\right>$ as a function of $q$
for different initial separations $R_0=1.7\eta$ (upper plot), $R_0=6.76\eta$
(middle plot), $R_0=70\eta$ (lower plot). The different curves represent
different times. Data from 3D simulations.}
\label{fig:7}
\end{figure}

In Fig. \ref{fig:7}, $\ln\left<\left(R(t)/ R_0\right)^q\right>$ as a
function of $q$ is presented for three different initial separations and for
various times. A clear crossing of zero at $q=-3$ can be observed up to time
$t=8\tau_{\eta}$ for $R_0=1.7\eta$. For $R_0=6.7\eta$, which lies
at the transition from the dissipative range, the initial crossing is shifted
above $-3$ (see intermediate Figure) a trend which is much more
pronounced for $R_0=70\eta$ at the inertial range.
This is the same trend that appeared in the 2D simulation and can be explained in a similar way.
Evidently, this shifted crossing point cannot also correspond to a conservation law, as it is time dependent, for both separations.
%%%%%%%%%%%%%%%%%%%%%%%%%%%%%%%%%%%%%%%%%%%%%%%%
%%%%%%%%%%%%%%%%%%%%%%%%%%%%%%%%%%%%%%%%%%%%%%%%

\begin{figure}
\includegraphics[width=1\linewidth]{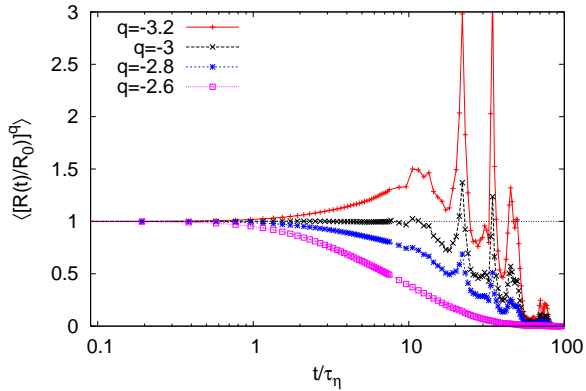}
\centering
\caption{$\left<\left(R(t)/ R_0\right)^q\right>$ as a function of time for
various $q$, the initial separation is $R_0=1.7\eta$. 
Different symbols represent $q=-2.6$, $q=-2.8$, $q=-3.0$ and $q=-3.2$
(from bottom to top). Data from 3D simulations.}
\label{fig:8}
\end{figure}

\begin{figure}
\includegraphics[width=1\linewidth]{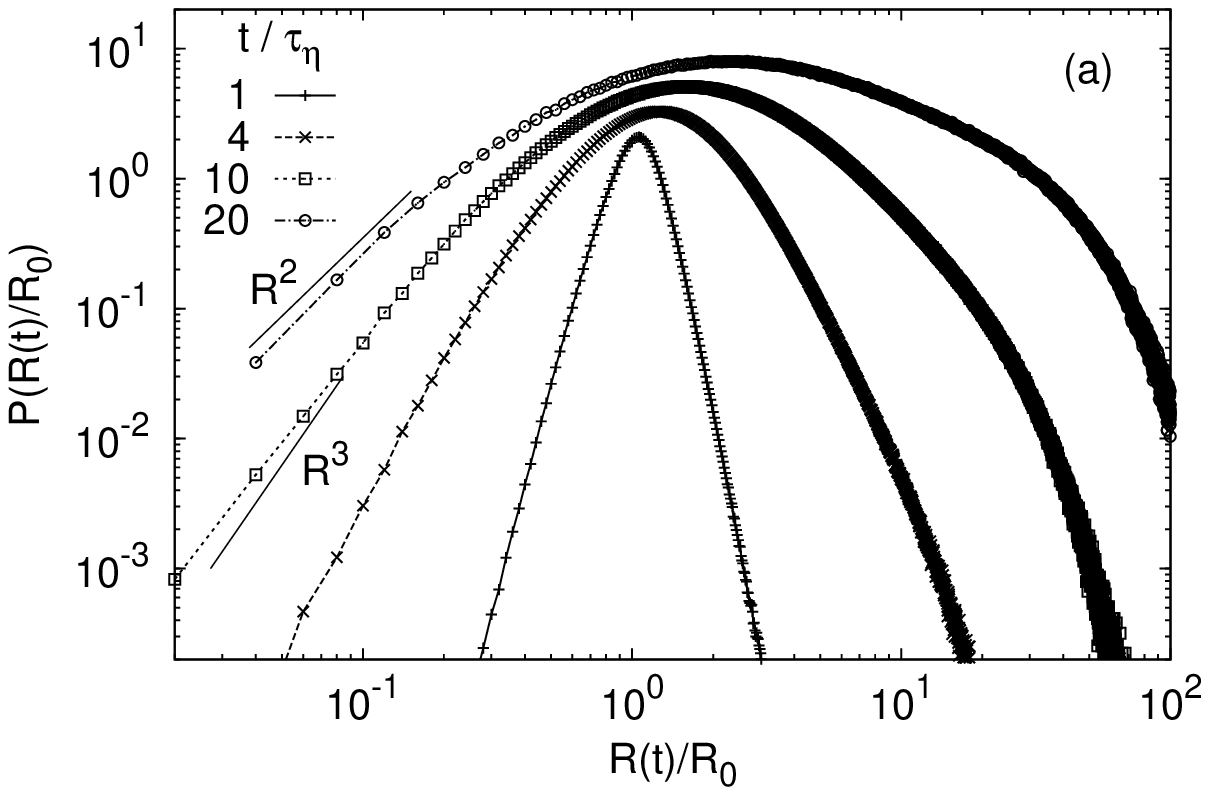}
\includegraphics[width=1\linewidth]{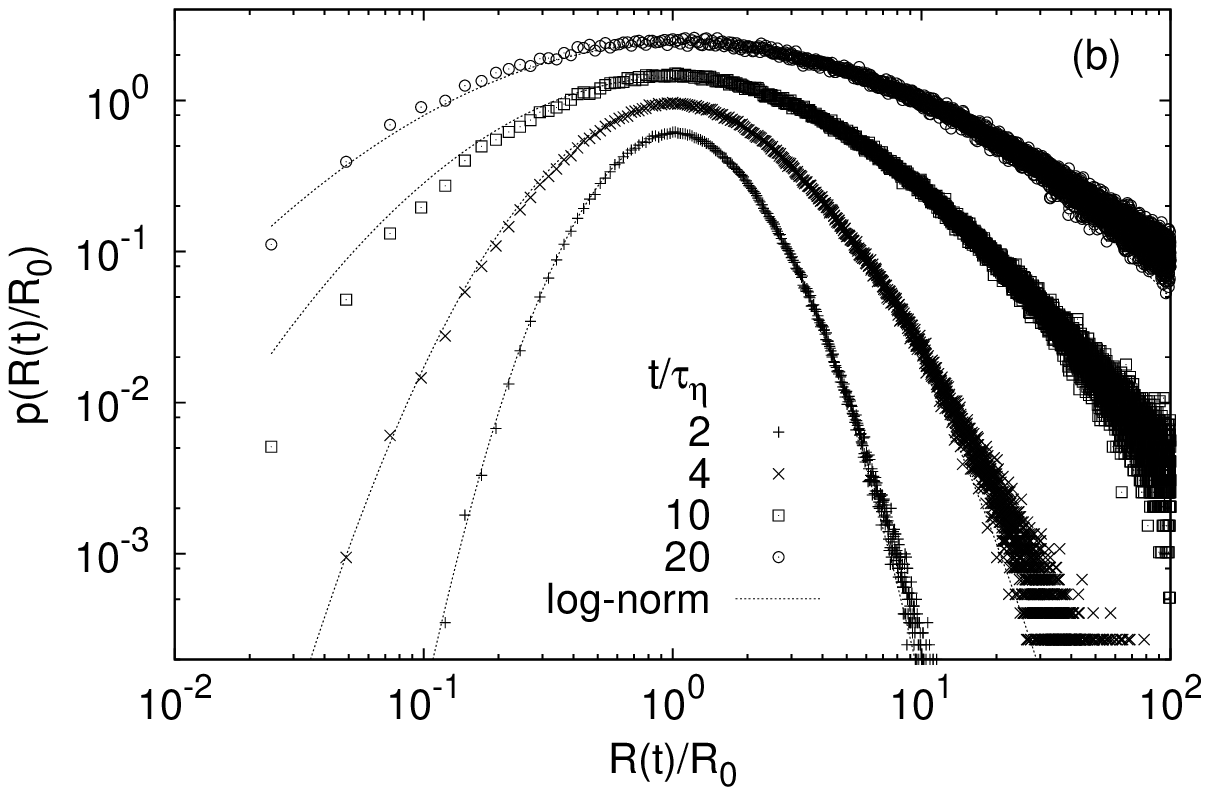}
\centering
\caption{$P(R(t)/R_0)$ for the smallest available initial separations in the 3D (a) and the 2D simulations (b) respectively.
While the 3D case shows the development of power-law tails for small
separations, in the 2D case all PDFs are close to log-normal curves, marked by
solid lines in (b)}.
\label{fig:9}
\end{figure}

Although theoretically $\left<\left(R(t)/ R_0\right)^{-3}\right>$ is an all time conservation law, in reality after some time pairs reach separations where a linear flow approximation no longer works, see also Fig. \ref{fig:2} and its discussion for the 2D flow.
We study the exit from the conservation regime of $\left<\left(R(t)/
R_0\right)^{-3}\right>$ for $R_0=1.7\eta$ in Fig. \ref{fig:8}. After $t\sim
8-10\tau_{\eta}$ wild fluctuations, as well as a decrease in value, appear
for $\left<\left(R(t)/ R_0\right)^{q}\right>$ with $q<-2$. This behaviour can
be explained by the formation of a power-law dependence on separation in the
separations pdf at small $R$, see Fig. \ref{fig:9}, together with a
shift of the average separation to larger values. Such a dependence has been
shown to develop for initial separations both in the inertial
range and in the dissipative one in \cite{Bitane2013a}. In Fig. \ref{fig:9}(a)
$P(R/R_0)$ is shown for different times for the 3D simulations. Indeed the left
tail shows a power law dependence $R^\alpha$, with $\alpha$ decreasing until
reaching $\alpha\sim 2$ at times $10\lesssim t/\tau \lesssim 20$. This explains
why the moment $\langle R^{-3}\rangle$ starts diverging somewhere in this
interval. For larger times we observe a further decrease in $\alpha$ (not
shown), compatible with the values observed in \cite{Bitane2013a}. This power law behaviour may be due to the contribution of trajectories which left the viscous subrange and returned to $R<R_0$. Indeed, $P(R/R_0)$ is
expected (at long enough times) to have a log-normal shape near its maximum, as long as the
separation remains in the viscous range \cite{Falkovich2001a}. In our 3D simulations,
$P(R/R_0)$ never displays a log-normal shape, probably as a consequence of the
limited extension of the viscous range. On the contrary, the PDF's of
separations are clearly log-normal in the 2D case Fig. \ref{fig:9}(b). Although
the statistics are not sufficient to distinguish the development of a power-law dependence in the
left tail in the 2D data, the large fluctuations in Fig. \ref{fig:2} point
towards a behaviour analogous to the 3D case.

%However, note that at time $t \approx 9\tau_{\eta}$ the average separation
%between pairs is $\sim 8\eta$.
%%%%%%%%%%%%%%%%%%%%%%%%%%%%%%%%%%%%%%%%%%%%%%%%
%%%%%%%%%%%%%%%%%%%%%%%%%%%%%%%%%%%%%%%%%%%%%%%%

%%%%%%%%%%%%%%%%%%%%%%%%%%%%%%%%%%%%%%%%%%%%%%%%
%%%%%%%%%%%%%%%%%%%%%%%%%%%%%%%%%%%%%%%%%%%%%%%%

%%%%%%%%%%%%%%%%%%%%%%%%%%%%%%%%%%%%%%%%%%%%%%%%%%%%%
\subsection*{Conservation in 3D - experimental results}
In the previous sections we have presented supporting evidence from 
numerical simulations, both in two and three dimensions, for the 
conservation of $\left< R^{-d}\right>$ for an isotropic flow at small 
enough scales. It is interesting to check if similar results can be
obtained from laboratory experimental data where the access to very 
small initial separation is limited by both physical and statistical 
constraints.
It is to this end that we turn to data from the experimental set up described
in \cite{Liberzon2012208}, where neutrally buoyant particles are tracked inside a
water tank of dimensions $32 \times 32 \times 50cm^3$. The turbulent flow, with
$Re_{\lambda}=84$ for our data set, is generated by eight propellers at
the tanks corners and particles are tracked in space and time using four CCD
cameras.
The cameras are focused on a small volume of $1cm^3$ to resolve the smallest
scales (the Kolmogorov scale is $\eta=0.4mm$).
At such small scales, the flow is expected to be isotropic. We study pairs of
particles with initial separations of a few $\eta$, for which numerical and
previous experimental studies of turbulent flows found an approximately linear
flow \cite{Watanabe2007,Bewley2013,Zhou2000,Benzi1995,Ishihara2009,Lohse1995a}.

As isotropy and a linear dependence of velocity differences on separation are
the two assumptions entering the prediction of the conservation of $\left<
R^{-3}\right>$, the experimental set up described above appears to be
appropriate to test it. Unfortunately, a direct check of these assumptions does
not seem to support their applicability.

%%%%%%%%%%%%%%%%%%%%%%%%%%%%%%%%%%%%%%%%%%%%%%%%
\begin{figure}
\includegraphics[width=1\linewidth]{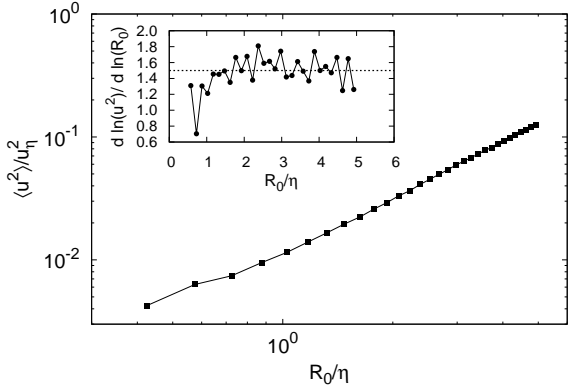}
\centering
\caption{Second order velocity structure function in the 3D experiment as
deduced from the initial relative velocity of pairs of particles with initial
separation $R_0$, normalized with the squared Kolmogorov velocity $u^2_\eta$. Inset: The local
slope $d\ln \left<\textbf{u}^2\right>/d\ln R_0$ as a function of the
separation. The dashed line represents the scaling  $\left<\textbf{u}^2\right>
\propto R_0^{1.5}$.}
\label{fig:10}
\end{figure}
%%%%%%%%%%%%%%%%%%%%%%%%%%%%%%%%%%%%%%%%%%%%%%%%
In Fig. ~\ref{fig:10} we present $\left<\textbf{u}^2\right>$, using the
relative velocity of particles with separation $R_0$. Here and in the following
$\left<\right>$
denotes an average over pairs in an ensemble with a given initial
separation\footnote{We thank Eldad Afik for suggesting his method of pair
selection for the ensembles \cite{Afik2014}, which we use in this work. We further filter out pairs of particles with a lifetime smaller than $0.5\tau_{\eta}$ to prevent contamination of the time evolution by a change of the particle ensemble}.
When it is applied to functions of the relative velocity or acceleration at a
given scale it denotes the average over the pairs with the corresponding
initial separation, taken at $t=0$.
As can be seen from the inset of Fig. \ref{fig:10}, if any scaling regime,
$\left<\textbf{u}^2\right>\propto R_0^{\zeta_2}$, is to be assigned, it would
be with scaling exponent $\zeta_2=1.5$ rather than $\zeta_2=2$. It seems that
in order to access the linear flow regime in this system even smaller distances
should be probed, which is not experimentally accessible at the moment.

For separations at the transition between inertial and dissipative ranges, as
those apparently probed in this experiment, there is currently no theoretical
prediction regarding the existence of a conservation law. Still, in the spirit
of the conservation law in the dissipative range, we can look for $q$ such that
$\left<\left(R(t)/R_0\right)^{q}\right>=1$.
Notice that as  a function of $q$, $\left<\left(R(t)/R_0\right)^{q}\right>$ is
convex and at each time $t$ it crosses $1$ at $q=0$ and either has one more
crossing point, denoted by $q^*(t)$, or none.
It would be possible to establish the existence of a conservation law, with
$q=q^*$, if $q^*$ exists and is time independent.
To get an idea of what value $q^*$ can take we use the Taylor expansion of
$\left<R^q(t)\right>$ around $t=0$ (no assumption is made about the nature of the flow):

\begin{equation}
\label{Rnstar}
\left< \left(\frac{R(t)}{R_0}\right)^q\right> =1+\frac{\left< u_l^2 \right>}{ R_0^2}\frac{qt^2}{2}\left[ q-q_e(t)\right]+O(t^3)
\end{equation}
with
\begin{align}
q_e(t)\equiv c-\frac{2\left<u_l\right> R_0}{t \left<u_l^2\right>}; && c\equiv 2-\frac{\left<\textbf{u}^2\right>+\left<a_l\right> R_0}{\left<u_l^2\right>}
\label{nstar}
\end{align}
where $u_l=\textbf{u}\cdot \hat{R}_0$, $a_l=\textbf{a}\cdot \hat{R}_0$ and $\textbf{a}$ is the relative acceleration. This implies that
the conservation law described above can exist only if $q_e(t)$ is time independent, so that $q^*=q_e=c$. This is of course only a necessary but not a sufficient condition.

The time dependence of $q_e$ comes from the first order in the Taylor expansion $\propto \left<u_l\right>$. It, as  well as $\left<a_l\right>$, should be zero for a statistically homogeneous or a statistically isotropic flow. Obviously in an experiment $\left<u_l\right>$ can never be exactly zero, but it can be small enough so that at the shortest times measured the time dependent term in $q_e$ would be negligible compared to the time independent part, $c$ in (\ref{nstar}).
%%%%%%%%%%%%%%%%%%%%%%%%%%%%%%%%%%%%%%%%%%%%%%%%%%%%%%%%%%%%%%%%%%%%%%
\begin{figure}
   \includegraphics[width=1\linewidth]{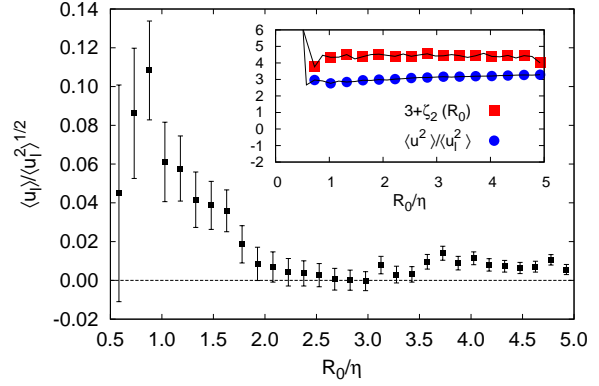}
   \centering
    \caption{Average longitudinal relative velocity as a function of initial separation for pairs of particles at $t=0$. The velocity is normalized by the initial rms longitudinal relative velocity. Inset: The LHS and the RHS of equation (\ref{zeta2}) are displayed as a function of $R_0$.}
    \label{fig:11}
\end{figure}
%%%%%%%%%%%%%%%%%%%%%%%%%%%%%%%%%%%%%%%%%%%%%%%%%%%%%%%%%%%%%%%%%%%%%%
In the inset of Fig. \ref{fig:11} we present $\left<u_l\right>$ normalised by the rms relative velocity. The strong bias towards positive $\left<u_l\right>$ for $R_0<2\eta$ is probably the result of the filtration of pairs with life times smaller than $0.5\tau_{\eta}$. Indeed, pairs with approaching particles at small separations are harder to track and are therefore frequently short lived as well as more prone to errors. A similar problem might also be the cause of the positive bias for $R_0\gtrsim 3\eta$, where particles with a large relative velocity are more common. Then, particles approaching each other with a large speed reach small scales quickly and are lost more easily. When the restriction on the life time of the pairs is lifted a negative bias emerges. It therefore appears that using $\left<u_l\right>$ for pairs of particles to determine the isotropy or homogeneity of the flow is problematic.
For our purposes, note that for an observation of the conservation of $\left< R^c(t)\right>$ to be possible the value of $\left<u_l\right>$ must be as small as possible. Initial separations $R_0=2.6\eta-3\eta$ seem most appropriate.

For a d-dimensional incompressible flow that is statistically isotropic a simpler formula for $q_e$ can be written. For such a flow (see also \cite{VanDeWater1999})
\begin{equation}
\left\langle \textbf{u}^2 \right>-\left\langle u_l^2 \right>= \frac{1}{R_0} \frac{d}{dR_0} \left(R_0^2 \left\langle u_l^2\right> \right).
\label{iso}
\end{equation}
which can be used to write
\begin{align}
\frac{\left\langle \textbf{u}^2 \right>}{\left<u_l^2\right>} =(d+\zeta_2(R_0)); && \zeta_2(R_0)=\frac{ d\ln \left<u_l^2\right>}{d\ln R_0}
\label{zeta2}
\end{align}
implying that in (\ref{nstar})
\begin{equation}
q_e=c=2-d-\zeta_2(R_0)=-1-\zeta_2(R_0).
\label{nstariso}
\end{equation}
As a side note, we remark that if $\zeta_2$ is constant, like in the inertial range, the initially slowest changing function, $R(t)^c=R(t)^{2-d-\zeta_2}$, is independent of $R_0$ \cite{Falkovich2013a}.
More generally, requiring only $\left<u_l\right>=0$, the $R_0$ independent slowest varying (among twice differentiable functions of $R$) function $F(R)$, is determined by demanding  $\left< d^2F(R)/dt^2\right>=0$, resulting in
\begin{equation}
\label{F(R)}
F(R)\propto\int_a^R  \exp\left[\int_a^x\frac{ -\left\langle u_{T}^2 \right>}{z  \left\langle u_l^2 \right>} dz\right] dx.
\end{equation}
 with $a$ an arbitrary separation and $u_{T}^2= \textbf{u}^2-u_l^2$.
For a 3-dimensional isotropic flow this formula reduces with the help of (\ref{iso}) to
\begin{equation}
\label{F(R)iso}
F(R)\propto \int_{a}^R \frac{1}{x^{2}\left\langle u_l^2 \right>} dx.
\end{equation}
Here, however, we focus on the simpler functions $F(R)=R^{q_e}$, which, through $q_e$, are $R_0$ dependent.

The isotropy assumption is checked using equation (\ref{zeta2}), as suggested in \cite{VanDeWater1999}, in the inset of Fig. \ref{fig:11} where we present both the LHS of this equation and its RHS as a function of initial separation. The difference between the two is about $1.5$ for most separations, implying that isotropy is violated.
It could also be that the flow is, in fact, isotropic: although in theory  measuring relative velocities for pairs of particles at a given distance is the same as using two fixed probes in the flow, in practice it might not be so  (for example due to correlations entering through the method of pair selection). 

%%%%%%%%%%%%%%%%%%%%%%%%%%%%%%%%%%%%%%%%%%%%%%%%%%%%%%%%%%%%
\begin{figure}
    \includegraphics[width=1\linewidth]{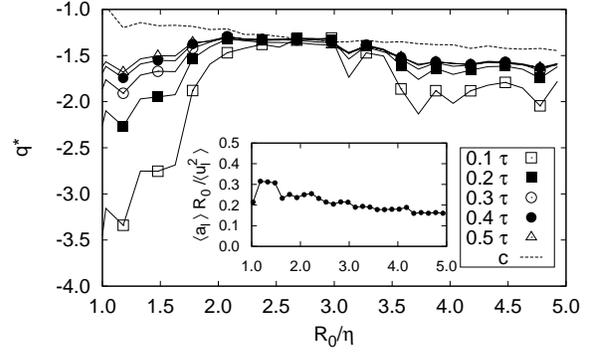}
    \centering
    \caption{The exponent $q^*(t)$, deduced by matching $\left< R(t)^{q^*}\right>=R_0^{q^*}$, plotted as a function of $R_0$ at different times. The dashed line represent the prediction $q^*(t)=c$, with $c$ from (\ref{nstar}). Inset: The negative of the contribution to $c$ due to $\left<a_l\right>$, $R_0 \left<a_l\right>/\left<u_l^2\right>$, plotted as a function of $R_0$.}
    \label{fig:12}
\end{figure}
%%%%%%%%%%%%%%%%%%%%%%%%%%%%%%%%%%%%%%%%%%%%%%%%%%%%%%%%%%%%%%%
In any case, we conclude that the prediction (\ref{nstariso}) should not work for our data, as indeed it does not, and instead use in the following the more general definition of $c$, (\ref{nstar}).
In Fig. \ref{fig:12} we present the exponent $q^*(t)$, as deduced from the data, for five consecutive times and compare it to $c$.
As expected, $q^*(t)$ appears approximately constant and equal to $c$ for the separations where $\left<u_l\right>$ is smallest. Note that $c$ has a non-negligible contribution coming from the deviation of $\left<a_l\right>$ from zero. This contribution is displayed in the inset of Fig. \ref{fig:12}.
%%%%%%%%%%%%%%%%%%%%%%%%%%%%%%%%%%%%%%%%%%%%%%%%%%%%%%%%%%%%%%%
\begin{figure}
    \includegraphics[width=\linewidth]{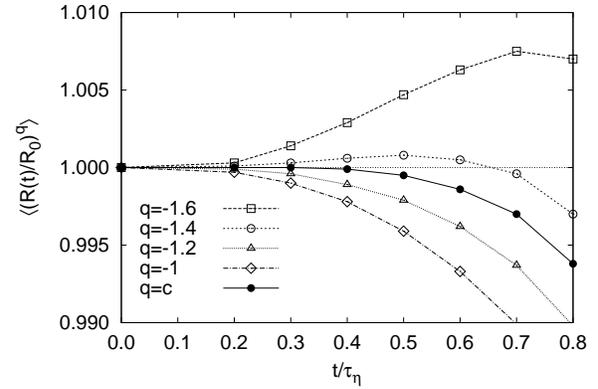}
    \centering
    \caption{The evolution in time of $\left<\left(R(t)/ R_0\right)^q\right>$ for pairs with initial separation of $R_0=2.6\eta$. The dashed dotted curve represents $q=c=-1.32$}
    \label{fig:13}
\end{figure}
%%%%%%%%%%%%%%%%%%%%%%%%%%%%%%%%%%%%%%%%%%%%%%%%%%%%%%%%%%%%%%%

We are now in a position to ask if for the separation where initially $q^*(t)\approx c$, $\left< R^{c}(t)\right>$ is just a slowly varying function or a true conservation law. In Fig. \ref{fig:13}, for the separation $R_0=2.6\eta$, we compare $\left< \left(R(t)/R_0\right)^{q}\right>$ as a function of time for different $q$s, including $q=c$. After a time shorter than $\tau_{\eta}$, $\left< R^{c}(t)\right>$ noticeably decreases. During this time the average separation between particles in a pair changes by less than one percent.
It is thus improbable that $\left< R^{c}(t)\right>$ is truly constant up to that time, but rather that it changes slowly, meaning it is not a conserved quantity.
We conclude that, at the scales accessible to this experiment, a power-law-type of conservation law probably does not exist but a slowly varying function of the separation can be identified.

\textbf{Acknowledgements} A.F is grateful to G. Falkovich and E. Afik for many useful discussions and comments. She would also like to thank J. Jucha for her valuable input on the experimental data and O. Hirschberg for recognizing the fluctuation relations. A.F is supported by the Adams Fellowship Program of the Israel Academy of Sciences and Humanities.

%\setcitestyle{numbers}
%\bibliography{MyCollection}	

\end{document}